\documentclass[12pt]{iopart}
\usepackage{graphicx}
\usepackage{lscape}
\usepackage{arydshln}

\begin{document}

\title{Phase diagram of hole-doped cuprates based on $^{17}$O and $^{63}$Cu NMR quadrupole splittings}

\author{Damian Rybicki$^{1,2}$, Michael Jurkutat$^1$, Steven Reichardt$^1$, J\"urgen Haase$^1$}

\address{$^1$ University of Leipzig, Faculty of Physics and Earth Sciences, Linnestr. 5, 04103 Leipzig, Germany}
\address{$^2$  AGH University of Science and Technology, Faculty of Physics and Applied Computer Science, Department of Solid State Physics, al. A. Mickiewicza 30, 30-059 Krak\'{o}w, Poland}
\ead{rybicki@physik.uni-leipzig.de}

\begin{abstract}
The phase diagram of the superconducting cuprates is often used to show how their electronic properties change as a function of the mean doping level, i.e., the average hole content of the CuO$_2$ plane.  In Nuclear Magnetic Resonance (NMR) experiments average doping, as well as the distribution of these holes between planar Cu and O reveals itself through the quadrupole splittings of the $^{63,65}$Cu and $^{17}$O NMR. Here we argue based on all published NMR data available to us in favor a new type of phase diagram that has the planar oxygen quadrupole splitting and with it the planar oxygen hole content as abscissa rather than the average hole content of the CuO$_2$ plane. In such a plot the superconducting domes of the different cuprate families are shifted horizontally according to their maximum critical temperature $T_{\rm c,max}$ set by the chemistry of the parent material, which determines its oxygen hole content. The higher the O hole content the higher $T_{\rm c,max}$ that can be achieved by actual doping. These findings also offer a strategy for finding cuprates with higher $T_{\rm c,max}$.
\end{abstract}

\pacs{74.25.nj, 74.72.Gh}

\submitto{\JPCM}

\maketitle

\section{Introduction}
The well-known phase diagram of the hole doped high-temperature superconducting cuprates (HTSCs) has  at zero doping ($x=0$) the antiferromagnetic insulator and by increasing $x$ this state disappears rapidly. The so-called "superconducting dome" appears at a few percent of doping, with the superconducting transition temperature $T_{\rm c}$ reaching its maximum ($T_{\rm c,max}$) near $x\approx0.15$ (optimal doping). Further increase of the doping causes a gradual decrease of $T_{\rm c}$ on the overdoped side. Why $T_{\rm c,max}$ is so different for different HTSC is not understood. It is believed that the understanding of the various electronic phases that appear to be generic to the different regions of the phase diagram will hold the clue also for understanding $T_{\rm c,max}$. This believe is not supported by our results discussed below. We will argue, based on all published NMR data available to us, that doping the HTSC only unlocks the $T_{\rm c,max}$ that is strongly influenced by the parent's hole sharing between planar copper and oxygen. A phase diagram that has the planar oxygen hole content as abscissa, rather than $x$ leads to well-ordered superconducting domes according to their $T_{\rm c,max}$.

NMR is a powerful local probe that contributed tremendously to the understanding of the chemical and electronic structure of the HTSCs\cite{Slichter2007}. For example, NMR can measure the electronic spin susceptibility with shift and relaxation measurements, and NMR discovered early on the pseudo-gap above  $T_{\rm c}$ on the underdoped side of the phase diagram, a property markedly different from those of classical superconductors.\cite{Hebel1957} For many years, it was believed that the shift and relaxation measurements can be understood with a single electronic fluid's spin response. However, with a set of NMR shift experiments at ambient and very high pressures on various systems it was shown, recently, that a single temperature dependent electronic spin component cannot explain the data.\cite{Haase2009, Haase2012, Meissner2011} Here, we do not discuss the magnetic properties of the HTSC, but address the quadrupole splitting of the NMR of planar Cu and O.

Since $^{63,65}$Cu and $^{17}$O possess an electric quadrupole moment (nuclear spin $I>1/2$), the quadrupole interaction gives additional insight into structural details of the HTSCs. In the absence of a magnetic field, the quadrupole interaction splits the degenerate nuclear spin levels, and in a high magnetic field, i.e., where the Zeeman interaction dominates the quadrupole interaction, the latter shifts the levels so that the NMR lines split into $2I$ transitions.\cite{Slichter1990} The splitting depends on the orientation of the magnetic field with respect to the crystal axes, and from angular dependent studies one can determine the principle axes values and the orientation of the tensor of the electric field gradient (EFG) that interacts with a nucleus' quadrupole moment. 

For the electron doped systems there are only few $^{63,65}$Cu NMR studies known, and $^{17}$O data are lacking, completely. Luckily, the situation is quite different for the hole-doped HTSCs where a large number of studies is available. These show that the quadrupole splittings for both nuclei in the CuO$_2$ plane depend linearly on the average doping level and thus the total hole content of the CuO$_2$ plane, and various models including first-principle calculations addressed this observation, e.g. Ref. ~\cite{zheng_local_1995, Asayama1996, haase_planar_2004}. 

Here we investigate all  published  $^{17}$O and $^{63}$Cu NMR data accessible to us (collected in Tabs.~\ref{tab:one} and \ref{tab:two}), and without invoking a special quantitative microscopic model that relates the NMR splitting to the hole content of the various orbitals we arrive at important conclusions. We show that the splittings at both nuclei are very good measures of the doping level, and we argue that different parent materials ($x=0$) differ in the hole distribution between Cu and O, only. It emerges that $T_{\rm c,max}$ is set by this distribution such that the bigger the planar O hole content the higher the $T_{\rm c,max}$. The actual doping only unlocks this potential (probably by destroying the Cu based magnetism). Thus, with respect to the number of O or Cu holes different families of HTSCs occupy very different regions in a phase diagram that has oxygen splitting and with it the number of oxygen holes as abscissa. Here, all HTSCs appear ordered, the higher the oxgygen hole content of the parent material the larger $T_{\rm c,max}$ of that system, irrespective of peculiarities often discussed, e.g., inhomogeneities.

\section{Results and Discussion}

First, we introduce some definitions and discuss some general properties that determine the quadrupole splittings in HTSCs. We begin with the parent materials and denote the EFG at a nucleus by $\Xi$ (we omit the label for O and Cu), a symmetric, traceless tensor with the three principal axis components $\xi_{ii}$, i.e.,
\begin{equation}
\Xi=\left(\xi_{11}, \xi_{22}, \xi_{33}\right)=\xi_{33}\left(-\frac{1}{2}(1-\eta_\xi), -\frac{1}{2}(1+\eta_\xi), 1\right).
\label{eq:xi}
\end{equation}
The asymmetry parameter $\eta_\xi$ is given by,
\begin{equation}
\eta_\xi=\frac{\xi_{11}-\xi_{22}}{\xi_{33}},
\end{equation}
with the usual definition that $\xi_{33}$ has the largest magnitude, i.e., $|\xi_{33}|\geq |\xi_{22}|\geq |\xi_{11}|$ ($\sum{\xi_{ii}}=0$).

Based on symmetry arguments, one expects for the parent compound an EFG at planar Cu (in square planar arrangement with planar O) that is axially symmetric ($\eta_\xi=0$). Since there is a substantial hole content in the Cu $3d(x^2-y^2)$ orbital, a large quadrupole splitting, i.e., a large $\xi_{33}$, is expected ($\xi_{22}=\xi_{11}=-\xi_{33}/2$; note that we cannot determine the sign of the $\xi_{ii}$, hence we take $\xi_{33}$ to be positive). One expects the principle axis of $\xi_{33}$ to coincide with the crystal c-axis, i.e., it is perpendicular to the CuO$_2$ plane. The charges in Cu 3$d(z^2-r^2)$ or $4p$ orbitals do not affect the local Cu symmetry ($4s$ will not contribute to the EFG, at all). The situation is expected to be different at planar O, which is in an almost full shell configuration (for a full shell the EFG is zero). Therefore, one expects a small quadrupole splitting with its largest principle component along the $2p_\sigma$ bond due to hybridization of Cu $3d(x^2-y^2)$ and O $2p_{\sigma}$ orbitals\cite{haase_planar_2004}. The asymmetry of the tensor is expected to be substantial since the two axes perpendicular to the O $2p_\sigma$ bond are not equivalent. 

Doping the parent compounds must change the Cu and/or O EFGs, but will not change the fundamental local symmetry at either nuclear site. 
In the most simple scenario one may assume that hole doping ($x$) creates an additional electric field with an axially symmetric EFG ($X$),
\begin{equation}
X=\left(x_{11}, x_{22}, x_{33}\right), \;\;\;\; X=x_{33}\left(-\frac{1}{2}, -\frac{1}{2}, 1\right).
\label{eq:x}
\end{equation}
For Cu this is mandatory, but for O the other $2p$ orbitals might be affected by the doping, as well. In the experiment we measure the sum of both EFGs ($\Theta=\Xi+X$), 
\begin{equation}
\Theta=\theta_{33}\left(-\frac{1}{2}(1-\eta_\theta), -\frac{1}{2}(1+\eta_\theta), 1\right), 
\end{equation}
and with (\ref{eq:x}) and (\ref{eq:xi}) we have,
\begin{eqnarray}
\theta_{11}=&-\frac{1}{2}\theta_{33}-\frac{1}{2}\xi_{33}\eta_\xi, \;\;\; \theta_{22}=-\frac{1}{2}\theta_{33}+\frac{1}{2}\xi_{33}\eta_\xi, \;\;\;
\theta_{33}=\xi_{33}+x_{33}\\
\eta_\theta=&\frac{\xi_{33}\eta_\xi}{\xi_{33}+x_{33}}.
\end{eqnarray}
One recognizes that, 
\begin{equation}
  \theta_{33}\eta_\theta=\xi_{33}\eta_\xi, \;\;\;\; \mathrm{i.e.,}\;\;\;\theta_{11}-\theta_{22}= \xi_{11}-\xi_{22},
\label{eq:product}
\end{equation}
and a symmetric contribution due to doping will not change the anisotropy given by the parent background EFG. Again, for Cu this is trivial and only $^{63}\theta_{33}$ will change with doping. For planar O one has to resort to the data collected in Tab.~\ref{tab:one}. They reveal two important things. Firstly, there is a steady increase of $^{17}\theta_{33}$ with doping, as well as an increase from one family to another if $T_{\rm c,max}$ of the family is also larger. Secondly, the anisotropy $^{17}\theta_{11}-^{17}\theta_{22}$ is rather similar for all HTSCs, and there appear to be only two groups of materials that differ in this background anisotropy at a finer level.

\begin{figure}
   \includegraphics[width=0.8\textwidth]{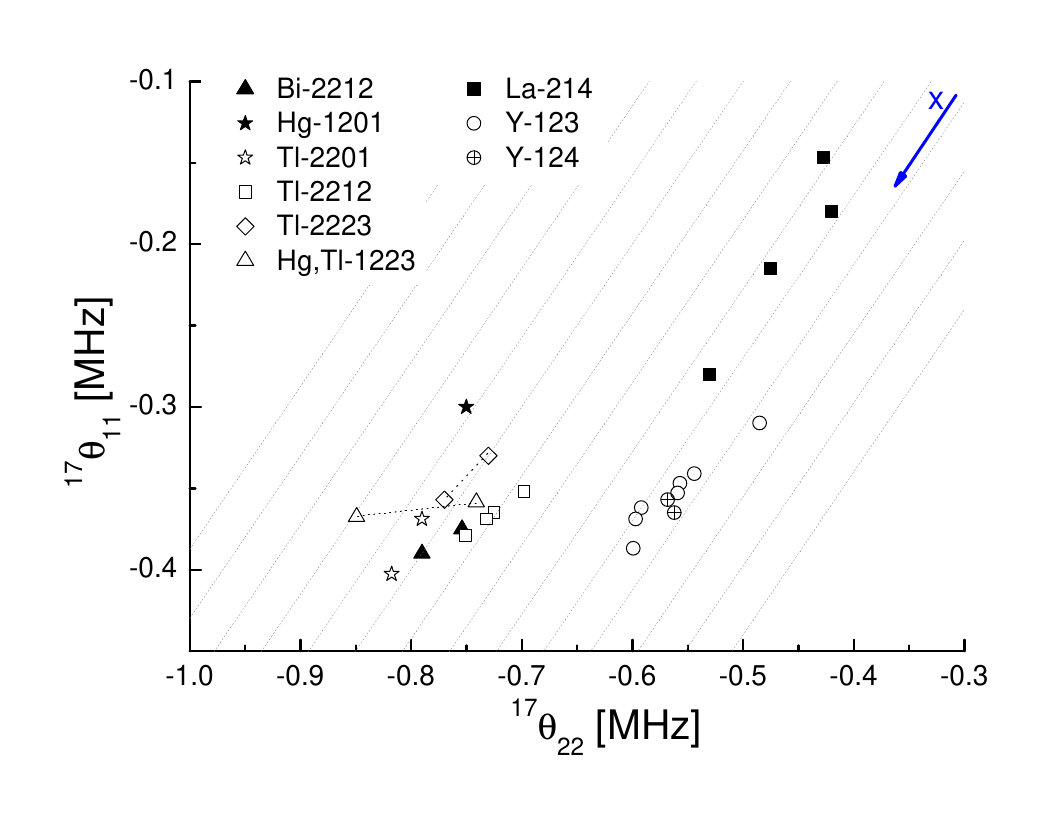}
   \caption{Components of the total planar $^{17}$O EFGs of different cuprate families and doping levels from Tab.~\ref{tab:one}: $^{17}\theta_{11}$ (along the crystal c-axis) vs.~$^{17}\theta_{22}$ (in the CuO$_2$ plane, perpendicular to O $2p_\sigma$). The average doping ($x$) is known to be linear in $^{17}\theta_{33}\left(=-^{17}\theta_{11}-^{17}\theta_{22}\right)$ and is indicated by the blue arrow. The data points for Tl-2223 and Hg,Tl-1223 that are connected by dashed lines belong to the two different O sites (inner and outer CuO$_2$ layer) for the same average doping\cite{Zheng1996}. The thin slanting dashed lines indicate a slope of 1, cf.~(\ref{eq:plot}).}
\label{fig:fig1}
\end{figure}

In order to investigate the $^{17}$O results further, we plot in Fig.~\ref{fig:fig1} the dependence $\theta_{11}(x)$ vs. $\theta_{22}(x)$ for the data shown in Tab.~\ref{tab:one}. We note that with (\ref{eq:product}) we have,
\begin{equation}
\theta_{11}(x)=\theta_{22}(x)+\xi_{33}\eta_\xi \equiv \theta_{22}(x)+(\xi_{11}-\xi_{22}).
\label{eq:plot}
\end{equation}
We recognize in Fig.~\ref{fig:fig1} the two groups of materials, easily. The dashed lines with slope one, cf. (\ref{eq:plot}), are guides to the eye. The doping dependence of La-214 fits the scenario that doping adds a symmetric tensor, only (this behavior is explained even quantitatively by the holes entering O 2$p_\sigma$ orbitals almost entirely \cite{haase_planar_2004}). Y-123 and Y-124 have nearly the same anisotropy, but deviate somewhat from the doping trend (perhaps since the other O orbitals are affected by the doping to some extent). Interestingly, the points for Y-123, Y-124 appear towards higher doping if one follows the line for La-214, i.e., further doping La-214 could generate the same points.  The second group of materials, appearing to the left in Fig.~\ref{fig:fig1}, seem to follow a similar trend as doping is concerned, but their anisotropy is clearly different. Note that these materials are much more anisotropic, but they also have the highest $T_{\rm c,max}$. To conclude, the data reveal that doping the different systems does not change the anisotropy substantially, and we can use $^{17}\theta_{33}$ as a good measure of the hole content at planar oxygen. This is in agreement with the well-established fact that $^{17}\theta_{33}$ is a linear function of $x$, but it suggests that the differences in $^{17}\theta_{33}$ between different families are largly due to a difference in the local planar oxygen hole content, as well.

\begin{figure}
   \includegraphics[width=0.7\textwidth]{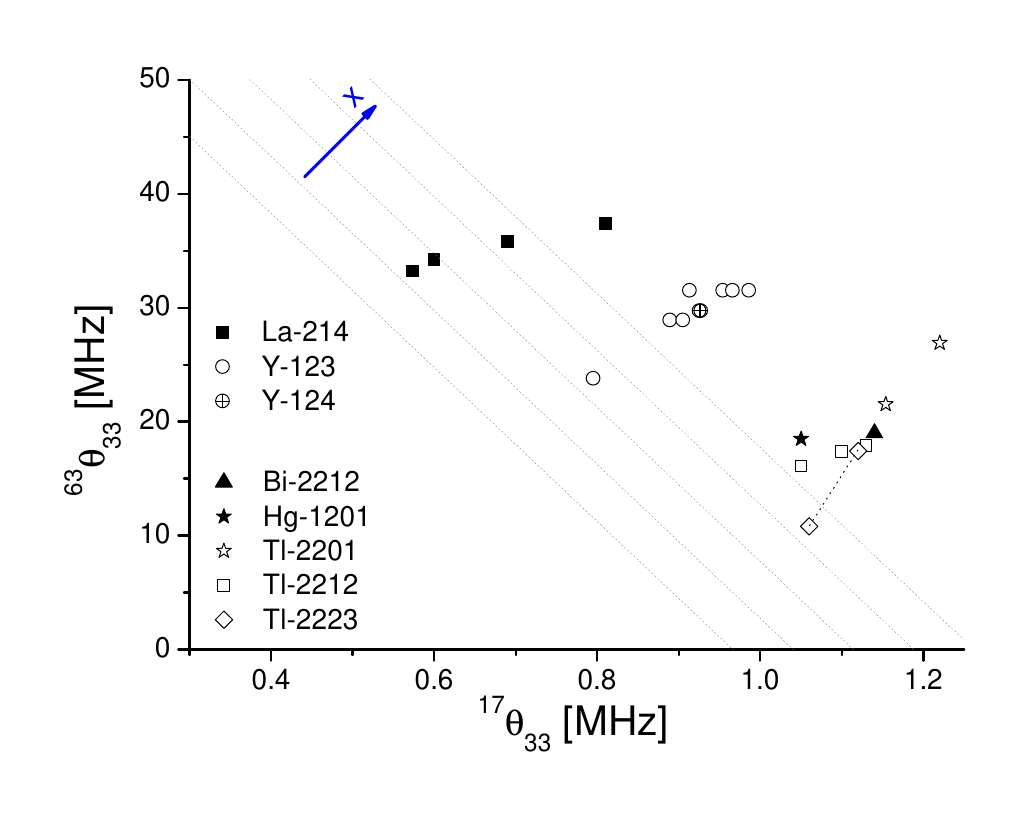}
   \caption{The largest EFG component at planar Cu ($^{63}\theta_{33}$) and O ($^{17}\theta_{33}$) plotted against each other (data from Tab.~\ref{tab:one} and \ref{tab:two}). The increase of the average doping is indicated by the blue arrow. The data points for Tl-2223 that are connected by a dashed line belong to the same doping level, but the two planar oxygen and copper sites from the inner and outer CuO$_2$ layers \cite{Zheng1996}. The slope of the thin, parallel lines follows from ref. \cite{haase_planar_2004} for fixed doping (e.g., $x$=const.).}
\label{fig:fig2}
\end{figure}

To further inquire about this conclusion, we plot in Fig.~\ref{fig:fig2} the largest principle components for Cu and  O against each other, i.e., $^{63}\theta_{33}$ vs.~$^{17}\theta_{33}$. We identify three groups of materials, now. First, we have La-214 with the largest $^{63}\theta_{33}$ and comparably small $^{17}\theta_{33}$ that strongly increases with doping (the holes enter the O 2$p_\sigma$ bond). The second group concerns the Y-123 data. The parent material of Y-123 starts at considerably lower $^{63}\theta_{33}$, but higher $^{17}\theta_{33}$; doping increases both paramters steadily. Interestingly, the stoichiometric Y-124 fits the set of data points. The third group of materials concerns those families that have the largest $T_{\rm c,max}$. Unfortunately, we do not have data for their parent materials, however, they must lie at much smaller $^{63}\theta_{33}$ and considerably larger $^{17}\theta_{33}$. Doping appears to increase both values, similar to what is observed for the other systems. 

Note, that a smaller Cu splitting ($^{63}\theta_{33}$) together with a larger O splitting ($^{17}\theta_{33}$) for a fixed total doping signals that the Cu hole has in part been transferred to planar O. In fact, we added a set of parallel lines from \cite{haase_planar_2004} representing constant doping. These lines support this scenario even quantitatively. Based on all the data one would argue that the planar O hole content is much higher in those systems that have the largest $T_{\rm {c,max}}$. 

\begin{figure}
   \includegraphics[width=0.6\textwidth]{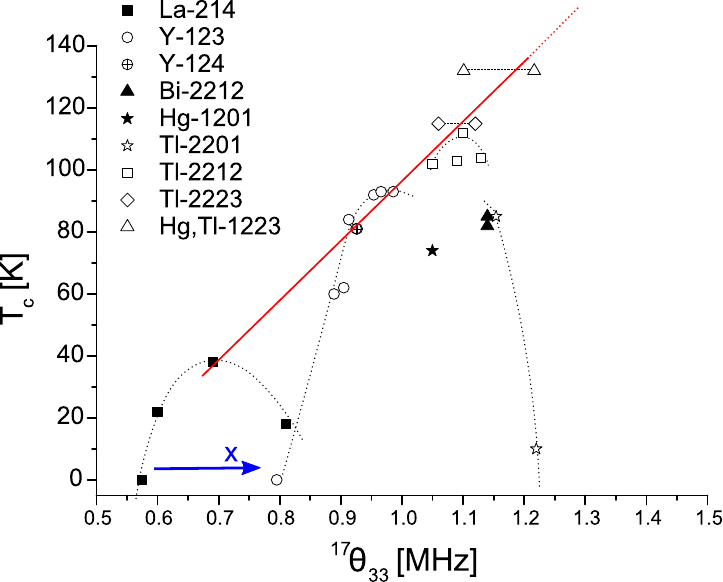}
   \caption{Phase diagram of the HTSCs based on $^{17}$O NMR splittings (data from Tab.~\ref{tab:one}). The black dotted lines are guides to the eye and connect different doping levels for one family; for Tl-2223 and Hg,Tl-1223 connected pairs belong to the same average doping, but the different planar O sites. The red line approximately connects $T_{\rm c, max}$. The blue arrow shows the increase of doping for all families.}
\label{fig:fig3}
\end{figure}

It is of course intriguing to plot the oxygen splitting  that represents the number of O $2p_\sigma$ holes against the critical temperature for all materials. This "phase diagram" is shown in Fig.~\ref{fig:fig3}. Due to the lack of data away from optimal doping for some materials we do not have the full parabolic behavior of $T_{\rm c}$ on $^{17}\theta_{33}$ for all systems. Contrary to the typical phase diagram of the cuprates Fig.~\ref{fig:fig3} suggests that there should be an offset between parabolas since the systems that can reach higher $T_{\rm c}$ do have a larger O $2p_\sigma$ hole content. We note that the values for $T_{\rm c, max}$ do scale with $^{17}\theta_{33}$. This is true irrespective of the kind of doping (and therefore of issues of inhomogeneity). Doping a particular family to achieve the highest $T_{\rm c}$ does increase the planar O hole content further, but starting with a parent material that already has a higher planar O hole content is even more important. Doping only unlocks the highest possible $T_{\rm c}$ (by destroying the Cu based magnetism).

\begin{figure}
   \includegraphics[width=0.6\textwidth]{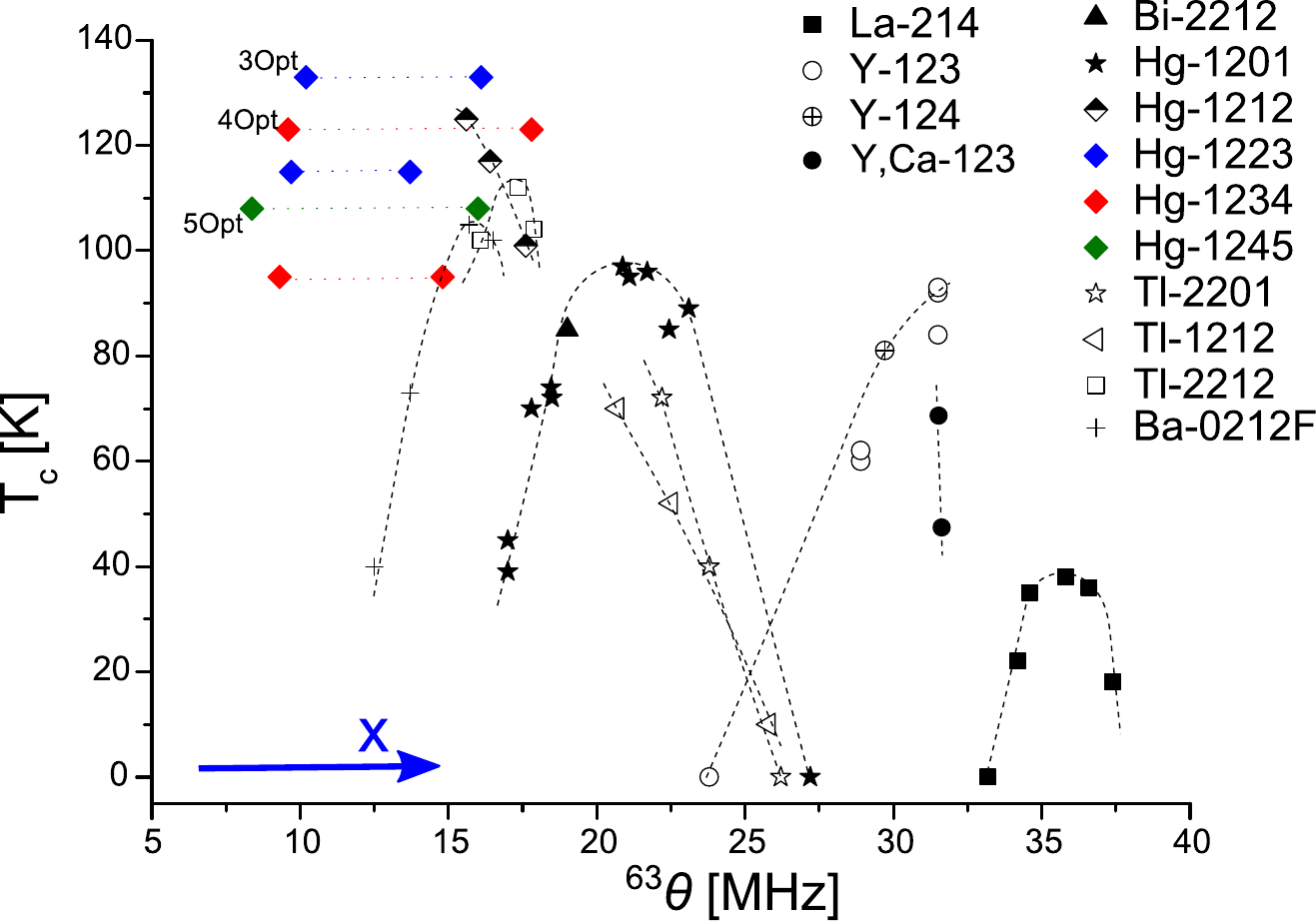}
   \caption{$T_{\rm c}$ versus the $^{63}$Cu NMR splitting (data from Tab.~\ref{tab:two}).  Black dashed lines are guides to the eye and connect different doping levels for one family (for the Hg- families with more than 3 CuO$_2$ layers the points are connected with colored dotted lines for the same average doping level). The blue arrow denotes the increase of the average doping for each family.}
\label{fig:fig4}
\end{figure}

For Cu the number of available data is much larger, cf.~Tab.~\ref{tab:two}, and we plot in Fig.~\ref{fig:fig4} $T_{\rm c}$ vs.~$^{63}\theta_{33}$ for the data we found in the literature. Clearly, there is the trend that materials with higher $T_{\rm c,max}$ will be found at lower $^{63}\theta_{33}$, but the trend is perhaps less striking compared to the planar O data. However, the Cu parabolas reveal more clearly the dependence on doping. The absolute changes in the Cu splitting are much bigger than for oxygen so that the resolution is better, however, one has to be cautious since $^{63}\theta_{33}$ depends also on the planar O hole content.\cite{haase_planar_2004}\\

To conclude, we have analyzed all data available to us for quadrupole splittings of planar Cu and O in the HTSCs. We find, as already stated in a number of publications, that the largest principle EFG components for Cu ($^{63}\theta_{33}$) and O ($^{17}\theta_{33}$) are linear functions of the average doping ($x$). We then conclude that even the differences in $^{63}\theta_{33}$ and $^{17}\theta_{33}$ between different families are by and large based on the differences in the Cu vs.~O hole content in the parent materials, i.e., for $x=0$. We then show that the highest critical temperature ($T_{\rm c,max}$) is a linear function of $^{17}\theta_{33}$, the planar oxygen hole content, i.e., the higher the oxygen hole content the higher $T_{\rm c,max}$. This leads us to proposing a new phase diagram for the HTSCs. It has the planar oxygen splitting (and thus the planar oxygen hole content) as abscissa, as opposed to just $x$. In our phase diagram, the superconducting domes of families with different $T_{\rm c,max}$ appear shifted, ordered with respect to the highest $T_{\rm c,max}$. While doping increases $T_{\rm c}$ it is not the key factor in determining $T_{\rm c,max}$. The proposed phase diagram might explain why one expects slight differences in the electronic properties for different families in the usual phase diagram.

\begin{landscape}
 \begin{table}
 \caption{Principal values of the planar oxygen EFG for various cuprates and doping levels ("und." - underdoped, "opt." - close to optimal doping, "ovd." - overdoped) . For Tl2223 and Hg,Tl-2223 there are two CuO$_2$ layers, the inner layer with O(1) and the outer layer O(2). Data are taken from: (a) Ref.\cite{haase_planar_2004}, (b) Ref.\cite{Mounce2013}, (c) Ref.\cite{Kambe1993}, (d) Ref.\cite{zheng_local_1995}, (e) Ref.\cite{Takigawa1994}, (f) Ref.\cite{Crocker2011}, (g) Ref.\cite{Trokiner1995}, (h) Ref.\cite{Gerashenko1999}, (i) Ref.\cite{Zheng1996}, (j) Ref.\cite{Lim1994}.
 \label{EFG-data}}
\small
  \begin{tabular}{| c | c | c | c | c | c | c | c|  }
Family	&	doping	&	T$_{\rm c}$		&		$^{17}\theta_{11}$	&	$^{17}\theta_{33}$	&	$^{17}\theta_{22}$	& $^{17}\eta$ &	$^{17}\theta_{11}$ - $^{17}\theta_{22}$\\
	&	&	[K]	& [MHz]	& [MHz]	& [MHz]	& [MHz]	&[MHz]	\\   \hline \hline
La-214$^a$	&	0	&	0	&		-0.147	&	0.574	&	-0.427	&	0.488	&	0.280	\\
  	&	0.075	&	22	&		-0.18	&	0.6	&	-0.42	&	0.400	&	0.240	\\
	&	0.15	&	38	&		-0.215	&	0.69	&	-0.475	&	0.377	&	0.260	\\
	&	0.24	&	18	&		-0.28	&	0.81	&	-0.53	&	0.309	&	0.250	\\   \hline
Y-123$^a$	&	0	&	0		&	-0.31	&	0.795	&	-0.485	&	0.220	&	0.175	\\
	&	0.6	&	60	&		-0.341	&	0.889	&	-0.544	&	0.228	&	0.203	\\
	&	0.63	&	62	&		-0.347	&	0.905	&	-0.557	&	0.232	&	0.210	\\
	&	0.8	&	84	&		-0.353	&	0.913	&	-0.559	&	0.226	&	0.206	\\
	&	0.96	&	92	&		-0.362	&	0.954	&	-0.592	&	0.241	&	0.230	\\
	&	1	&	93	&		-0.387	&	0.986	&	-0.599	&	0.215	&	0.212	\\
	&	1	&	93	&		-0.369	&	0.966	&	-0.597	&	0.236	&	0.228	\\  \hline
Y-124$^a$ 	&		&	81	&		-0.365	&	0.927	&	-0.562	&	0.213	&	0.197	\\
	&		&	81	&		-0.357	&	0.925	&	-0.568	&	0.228	&	0.211	\\  \hline
Hg-1201$^{b}$	&		&	74 (und)	&		-0.300	&	1.050	&	-0.750	&	0.429	&	0.450	\\
\hline																	
Tl-2201$^{c,d}$	&		&	85 (opt)	&		-0.369	&	1.154	&	-0.790	&	0.365	&	0.421	\\
	&		&	10 (ovd)	&	-0.403	&	1.220	&	-0.817	&	0.340	&	0.415	\\
\hline																	
Bi-2212$^{e,f}$	&		&	86 (opt)	&		-0.39 	&	1.14	&	-0.75	&	0.316	&	0.360	\\
                  &		&	82 (ovd)	&	     	-0.375	&	1.140	&	-0.754	&	0.332	&	0.379	\\
\hline
Tl-2212	&		&	103 (ovd)$^{g}$ 			&	-0.365 	&	1.09	&	-0.725	&	0.330	&	0.360	\\	
                              &		&	102 (und)$^{h}$ 		&		-0.352	&	1.05	&	-0.698	&	0.330	&	0.346	\\
                              &		&	112 (opt)$^{h}$ 		&		-0.369	&	1.10	&	-0.732	&	0.330	&	0.363	\\
                              &		&	104 (ovd)$^{h}$ 		&		-0.379	&	1.13	&	-0.751	&	0.330	&	0.372	\\  \hline
Tl-2223 O(1)	&		&	115 (ovd?)$^{i}$	&		-0.33	 &	1.06	&	-0.73 	&	0.377	&	0.400	\\ 
  O(2)	&		&		&		-0.357	&	1.12	&	-0.77 	&	0.369	&	0.413	\\ \hline
Hg,Tl-1223  O(1)	&		&	132 (opt)$^{j}$			&	-0.359 &	1.101 &	-0.741 	&	0.347	&	0.383	\\ 
 outer  O(2)	&		&				&	-0.368	&	1.217	&	-0.849 	&	0.396	&	0.482	\\ \hline

 \end{tabular}
\label{tab:one}
 \end{table}
\end{landscape}

 \begin{table}
 \caption{Principal values of the planar Cu EFGs for various cuprates and doping levels ("und." - underdoped, "opt." - close to optimal doping, "ovd." - overdoped) . For Tl- and Hg- families with more than 3 CuO$_2$ layers there are two Cu sites, the inner layer with Cu(1) and the outer layer Cu(2). Data are taken from: (a) Ref.\cite{haase_planar_2004}, (aa) Ref.\cite{Williams2001},  (b) Ref.\cite{Gippius1997}, (c) Ref.\cite{RybickiU}, (d) Ref.\cite{Rybicki2012b}, (e) Ref.\cite{Rybicki2009}, (f) Ref.\cite{Ohsugi1996}, (g) Ref.\cite{Horvatic1994}, (h) Ref.\cite{Julien1996a}, (i) Ref.\cite{Julien1996b}, (j) Ref.\cite{Magishi1996a}, (k) Ref.\cite{Breitzke2004}, (l) Ref.\cite{Itohara2010}, (m) Ref.\cite{Kotegawa2004},
 \label{EFG-data}}
\small
  \begin{tabular}{| c | c | c |  }
Family		&	T$_{\rm c}$		&	$^{63}\theta_{33}$	\\
	&		[K]	& [MHz]		\\   \hline \hline
La-214$^a$		&	0	&	33.2		\\
  	&		22	&	34.2	\\
	&		35	&	34.6			\\
	&		38	&	35.8		\\
	&		36	&	36.6	\\ 
	&		18	&	37.4	\\   \hline
Y-123$^a$	&		0	&	23.8		\\
	&		60	&	28.9		\\
	&		62	&	28.9		\\
	&		84	&	31.5			\\
	&		92	&	31.5		\\
	&		93	&	31.5	\\  \hline
Y-124$^a$ 			&	81	&	29.72	\\ \hline
Y,Ca-123	&		68 (ovd)$^{aa}$ 	&	31.55		\\
		&		48 (ovd)$^{aa}$  	&	31.65	\\   \hline
Hg-1201	&			39 (und)$^b$	&	17.0	\\ 
		&			45 (und)$^c$	&	17.0	\\
		&			70 (und)$^b$	&	17.8	\\ 
		&			72 (und)$^b$	&	18.5	\\
		&			74 (und)$^d$	&	18.46	\\
		&			95 (opt)$^b$	&	21.1	\\
		&			96 (opt)$^b$	&	21.7	\\
		&			97 (opt)$^e$	&	20.88	\\     
		&			89 (ovd)$^b$	&	23.1	\\
		&			85 (ovd)$^c$	&	22.4	\\
		&			0 (ovd)$^b$	&	27.2	\\   \hline
Hg-1212	&			125 (opt)$^f$	&	15.6	\\ 
		&			117 (ovd)$^f$	&	16.4	\\
		&			101 (ovd)$^g$	&	17.0	\\
		&			101 (ovd)$^h$	&	17.6	\\ \hline
Hg-1223 Cu(1)	&		115 (und)$^i$	&	9.7	\\ 
  Cu(2)	&				&	13.7		\\ \hdashline
 Cu(1)	&			133 (opt)$^j$	&	10.2	\\ 
  Cu(2)	&				&	16.1		\\ \hdashline
Hg,Cu-1223 Cu(1)			&	134 (opt)$^k$	&	15.3	\\ 
  Cu(2)	&				&	16.6		\\ \hline
Hg-1234 Cu(1)	&		85 (und)$^l$	&	9.3	\\ 
  Cu(2)	&				&	14.8		\\ \hdashline
Cu(1)	&			123 (opt)$^l$	&	9.6	\\ 
  Cu(2)	&				&	17.8		\\  \hline
Hg-1245 Cu(1)	&		108 (opt)$^m$	&	8.37	\\ 
  Cu(2)	&				&	16		\\ \hline
 \end{tabular}
\label{tab:two}
 \end{table}

  \begin{table}
 \caption{ctd of Tab.~\ref{tab:two}, (n) Ref.\cite{Ishida1994}, (o) Ref.\cite{Fujiwara1991}, (p) Ref.\cite{Magishi1996b},(r) Ref.\cite{Gerashenko1999}, (s) Ref.\cite{Han1994}, (t) Ref.\cite{Zheng1996}, (u) Ref.\cite{Shimizu2011}. }
\small
  \begin{tabular}{| c | c | c | }
Family	&	T$_{\rm c}$		&	$^{63}\theta_{33}$	\\
	&		[K]	& [MHz]		\\   \hline \hline
Bi-2212	&			86 (opt)$^n$	&	19.0 \\ \hline
Tl-2201	&			72 (ovd)$^o$	&	22.2\\ 
	&			40 (ovd)$^o$	&	23.8	\\
	&			0 (ovd)$^o$	&	26.2	\\ \hline	
Tl-1212	&			70 (ovd)$^p$	&	20.7\\ 
	&			52 (ovd)$^p$	&	22.5	\\
	&			10 (ovd)$^p$	&	25.8	\\ \hline														
Tl-2212	                		&	102 (und)$^{r}$ 		&	16.08     	\\
                              		&	112 (opt)$^{r}$ 		&	17.35    	\\
                              		&	104 (ovd)$^{r}$ 		&	17.87      	\\  \hline
Tl-2223 Cu(1)			&	125 (opt)$^{s}$	&	11.7		\\ 
  Cu(2)	&				&	16.4		\\ \hdashline
Cu(1)	&			115 (ovd?)$^{t}$	&	10.8		\\ 
  Cu(2)	&				&	17.4		\\ \hline
Ba-0212F	&			40 (und)$^u$	&	12.5\\ 
	&			73 (und)$^u$	&	13.7	\\
	&			105 (opt)$^u$	&	15.7	\\
	&			102 (ovd)$^u$	&	16.5	\\ \hline

 \end{tabular}
\label{tab:three}
 \end{table}

\section{Acknowledgments}
We would like to thank O.K. Andersen, A. Bussmann-Holder, G. V. M. Williams, A. Erb, Th. Meier, R. G\"uhne, and  O. P. Sushkov for helpful discussions. We also acknowledge financial support by Leipzig University, the DFG within the Graduate School Build-MoNa, the European Social Fund (ESF) and the Free State of Saxony. 

\newpage
\section*{References}
\bibliographystyle{unsrt.bst}
\bibliography{EFG}

\end{document}